\documentclass[epj]{svjour}
\usepackage{graphics}

\begin{document}
\title{Generic two-phase coexistence in nonequilibrium systems}
\author{M.A. Mu\~noz\inst{1} \and  F. de los Santos\inst{1} 
\and M. M. Telo da Gama\inst{2}
} 
\institute{Instituto Carlos I de F\'{\i}sica Te\'orica y Computacional,
Facultad de Ciencias, Universidad de Granada, 18071 Granada, Spain. 
\and Departamento de F{\'\i}sica da Faculdade de Ci{\^e}ncias e
Centro de F{\'\i}sica Te\'orica e Computacional da Universidade de
Lisboa,\\ Avenida Professor Gama Pinto, 2, P-1649-003 Lisboa Codex,
Portugal}
\date{Received: date / Revised version: date}
%
\abstract{
A beautifully simple model introduced a couple of decades ago, Toom's 
cellular automaton, revealed that non-equilibrium systems may exhibit 
generic bistability, i.e. two-phase coexistence over a finite area of 
the (two-dimensional) phase diagram, in violation of the equilibrium 
Gibbs phase rule. In this paper we analyse two interfacial models, 
describing more realistic situations, that share with Toom's model a 
phase diagram with a broad region of phase coexistence. 
An analysis of the interfacial models yields conditions for generic 
bistability in terms of physically relevant parameters that may be 
controlled experimentally.}

%
\maketitle

\section{Introduction}
\label{intro}

Gibbs' phase rule states that two-phase coexistence of a
single-component system, characterized by an $n$ - dimensional
parameter-space, may occur in an $n-1$ - dimensional region. For
example, the two equilibrium phases of the Ising model coexist on a
line in the temperature-magnetic-field phase diagram. Nonequilibrium
systems may violate this rule and several models, where phase
coexistence occurs over a finite ($n$-dimensional) region of the
parameter space, have been reported. The first example of this
behaviour was found in Toom's model \cite{Toom,Geoff,GG}, that
exhibits {\it generic bistability}, i.e. two-phase coexistence over a
finite region of its two-dimensional parameter space (see Section
\ref{sec:1}).  In addition to its interest as a genuine nonequilibrium
property, generic multistability, defined as a generalization of
bistability, is both of practical and theoretical relevance. In
particular, it has been used recently to argue that some complex
structures appearing in nature could be truly stable rather than
metastable (with important applications in theoretical biology), and
as the theoretical basis for an error-correction method in computer
science (see \cite{GG,Gacs} for an illuminating and pedagogical
discussion of these ideas).

The necessary ingredients to generate broad ($n$ - dimensional) 
phase-coexistence in Toom's model have been discussed in \cite{Geoff}, 
where general criteria were also identified. The main idea is that, 
to obtain generic bistability, a nonequilibrium mechanism that prevents  
the growth of the stable phase while enhancing the stability of the 
other one, must exist. This is achieved in Toom's model by introducing 
an explicit, somewhat artificial, asymmetry in the dynamic rules. 

Recent progress in nonequilibrium statistical mechanics has led to the 
analysis of more realistic, physically motivated, models that also exhibit 
generic bistability. In this paper we describe two such examples, both 
interfacial models of pinning-depinning transitions. The analogies and 
differences with Toom's model are described in detail providing the reader 
with a general view of the relevant nonequilibrium physical mechanisms. 
In particular, we identify the ingredients, of the physically 
motivated models, that are responsible for the emergence of generic 
multistability. 

This paper is organized as follows. In section \ref{sec:1} we review 
Toom's model. The following two sections are devoted to the description of 
the two interfacial models exhibiting broad phase-coexistence. A general 
discussion as well as the conclusions are given in the last section.

\section{Brief review of Toom's NEC model}
\label{sec:1}
In the following we describe very briefly Toom's NEC (North-East-Center) 
model and discuss some of its basic properties. More detailed descriptions 
may be found in the original papers by Toom \cite{Toom} and  
\cite{Geoff,GG,Geoff2}.

The model is defined on a two-dimensional square lattice, with sites 
occupied by a spin-like variable, $s_i = \pm 1$. The state of the  system
evolves by simultaneous updating of the spins according to the
following rules: {\bf i)} The value of any given spin is determined by 
the majority rule applied to a neighborhood that includes the spin
itself (C=center) and its neighbors to the north (N) and to the 
east (E). {\bf ii)} If as a result of (i) the spin points up (down), it is
inverted with probability $q$ ($p$). These rules are iterated leading 
eventually to a statistically stationary state. Different types of 
boundary conditions may be implemented but here we consider
periodic boundary conditions only.

\begin{figure}
\begin{center}
\resizebox{0.4\textwidth}{!}{%
\includegraphics{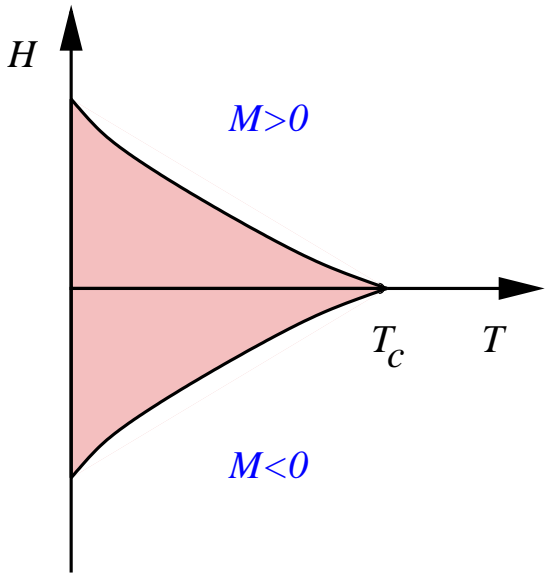}}
\caption{Phase diagram of Toom's model, showing the phases with
  positive and negative ``magnetization'' ($M$) as well as the broad
  coexistence region.}
\label{toom_pd}
\end{center}
\end{figure}

Toom's NEC model has two parameters, $q$ and $p$. Obviously, in the
symmetric situation $p=q$ both up and down spins are equally favored,
while for $p > q$ ($q > p$) up (down) spins are preferred. A bias
field $H$ may be defined, $H=(p-q)/(p+q)$, as the analog of the
external magnetic field in the standard Ising model. Likewise, the
noise intensity is measured by the temperature-like parameter
$T=p+q$. The associated phase diagram in the $(H,T)$ plane is shown in
figure 1. The solid lines are first-order phase boundaries between the
one-phase region, with positive or negative magnetization (unshaded
area), and the two-phase coexistence region (shaded area). The upper
and lower phase boundaries merge at a critical point at
$(H=0,T_c$). We note that the two-phase coexistence region in the
Ising model is the segment of line $T<T_c$ at $H=0$. By contrast, here
phases of positive and negative magnetization lose their stability at
$-H(T)$ and $H(T)$, respectively, and the coexistence surface is given
by $-H(T) < H < H(T)$ for $T<T_c$. In particular, the lower the
temperature, the larger the value of $|H|$ at the first-order phase
boundaries, implying that at low intensity of noise the system can
sustain phase coexistence even for relatively large values of the
bias. This is indeed surprising: In the presence of a relatively large
bias, a phase with the opposite magnetization may be truly stable.

Inspection of the dynamical rules leads immediately to the conclusion
that, in the deterministic limit $T=0$, an initially horizontal
interface, separating two semi-infinite planes of up and down spins,
will remain immobile. The same will happen if the interface is vertical, 
or oriented at $45^\circ$ along the NE-SW diagonal. On the other hand, if 
the interface is oriented at $-45^\circ$ along the NW-SE diagonal, it will 
move in a direction perpendicular to it, with constant velocity, and the 
phase to the right of the interface will advance downwards ``eroding'' 
the phase to the left. Thus, any island of, say, up spins in a sea of down 
spins is effectively eliminated since it can be inscribed in a right 
triangle with the hypotenuse along the NW-SE diagonal. If noise and 
bias are switched on, the situation does not change much: the phase to the
right of the interface will advance even if it is unfavored by the
bias, provided that both $T$ and $H$ are sufficiently small. Upon further
increase of the parameters, a point is reached where the unfavored 
phase is no longer stable, signaling the end of the broad coexistence 
region \cite{GG}.

The existence of a broad coexistence region in Toom's model has been
rigorously demonstrated \cite{Toom}. It was also shown that this feature 
does not depend on the discreteness of the spin variables or
of the space-time \cite{Geoff2}. The only relevant ingredient seems to 
be the spatial asymmetry of the nonequilibrium rules. These rules act 
differently on the interfacial motion depending on the initial 
orientation of the interface. As as a result, islands of spins of the 
minority phase are unstable even if the phase itself is favored by the 
bias field. 


As a final remark, we mention that the nature of the fluctuations
of NEC interfaces separating up- from down-spin regions, was 
investigated and was found to be related to {\it Kardar-Parisi-Zhang} 
(KPZ) \cite{Barabasi,HHZ} nonequilibrium dynamics in the biased case, 
and to (equilibrium) {\it Edwards-Wilkinson} (EW) relaxation dynamics
\cite{Barabasi,HHZ} in the unbiased one \cite{Toom-inter}.

\section{Nonequilibrium depinning of a bound interface}
\label{sec:2}

The second example to be discussed arises in the context of
nonequilibrium bound interfaces, and has relevance in nonequilibrium
wetting, synchronization transitions in extended systems, and general
pinning-depinning transitions
\cite{Lisboa,Marsili,Mallorca,Muller,synchro}.
Most of the material presented in this section is already known, but for
the  sake of completeness we include it here.

\begin{figure}
\begin{center}
\resizebox{0.4\textwidth}{!}{%
\includegraphics{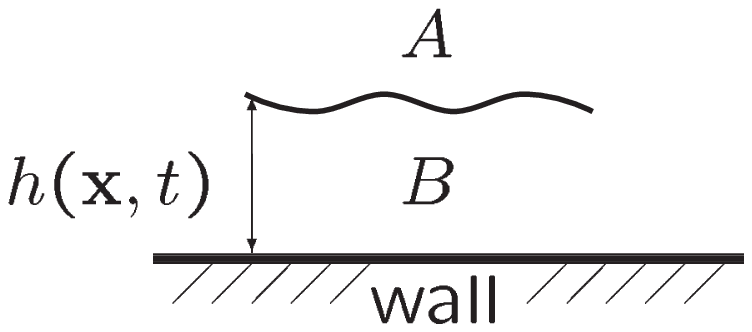}}
\caption{Interface between the $A$ and $B$ phases, at a 
distance from the wall}
\label{profile}
\end{center}
\end{figure}

\begin{figure}
\begin{center}
\resizebox{0.4\textwidth}{!}{%
\includegraphics{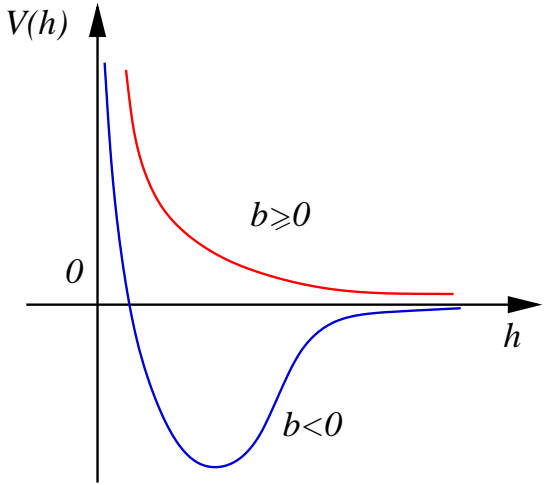}}
\caption{Binding potential for positive and negative values of $b$.
For $b<0$ the potential is attractive, exhibiting a local minimum
near the wall.}
\label{potential}
\end{center}
\end{figure}

\begin{figure}
\begin{center}
\resizebox{0.5\textwidth}{!}{%
\includegraphics{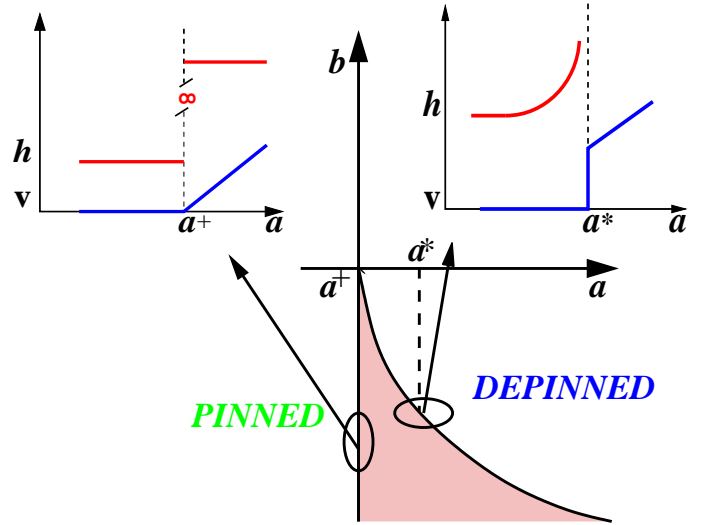}}
\caption{Phase diagram as a function of $a$ and $b$. A broad
phase coexistence region (shaded area) is observed. 
Insets: order-parameters (average velocity and average
height) at the phase boundaries delimiting the broad
coexistence region. In the leftmost (rightmost) panel the height
(velocity) changes discontinuously while the velocity (height)
changes continuously.}
\label{pd1}
\end{center}
\end{figure}

 Consider an interface separating two bulk phases, $A$ and $B$ (see
 figure \ref{profile}). Let $a$ be the chemical potential difference
 between these two phases, $D$ the surface tension of the $AB$
 interface, and $h({\bf x},t)$ the local height measured from a
 binding wall or substrate.  The interaction between the latter and
 the interface is usually modeled by a Morse potential
\begin{equation}
V(h)= b e^{-h}+ e^{-2h}/2,
\end{equation}
where the repulsive term restricts the interface 
from fluctuating into the unphysical region $h<0$, as shown  
in figure \ref{potential}. In the absence of conservation laws, the most
generic nonequilibrium interfacial equation is
\begin{equation}
\partial_t h({\bf x},t)= D \nabla^2 h + \lambda (\nabla h)^2+a
-{\partial V(h)\over \partial h} + \eta({\bf x},t),
\label{kpzww}
\end{equation}
where $\eta$ is a Gaussian white noise term, that accounts for the thermal
fluctuations, and $\lambda$ is the strength of the
KPZ non-linear term, which acts as an external
force pushing the tilted interfacial regions against the wall. Now,
two different physical situations may occur depending on the sign of $b$
\cite{MNreview,trends}:

For $b\ge0$, the wall is purely repulsive. As a function of $a$, a
continuous, nonequilibrium phase transition from a pinned to a moving
interface will occur. Here, we shall not consider this
transition. It has been studied extensively and the critical
exponents, scaling functions, etc, are well known (see
\cite{MNreview,trends} for recent reviews).
Assume then that $b$ is negative \cite{renorm}. 
In this case, the potential $V(h)$ includes an attractive term 
(see figure \ref{potential}) which describes
the affinity of the substrate for the $A$ phase \cite{Lisboa}. 
Such a term is also required in the context of
synchronization transitions \cite{synchro}.
Under these conditions, it can be argued that pinned and
depinned phases lose their stability at different points, provided
that $\lambda < 0$. For the depinned phase this happens at $a=a^+$ for any
value of $b$ (figure \ref{pd1}). To see this, notice that as far as
their long-time properties are concerned, depinned interfaces may be 
considered effectively free. Thus, the wall potential may be neglected 
implying that the average interfacial velocity $\langle v \rangle$ is 
independent of $b$; this implies in turn that the locus of the depinning
transition, where $\langle v \rangle$ changes from positive
(depinned) to zero (pinned), is also independent of $b$. Note that the 
transition is continuous in terms of the usual order-parameter, $\langle v 
\rangle$, but discontinuous in terms of the average interface position, 
$\langle h \rangle$, that jumps from infinity (depinned) to a finite 
value close to the potential minimum (pinned).

\begin{figure}
\begin{center}
\resizebox{0.5\textwidth}{!}{%
\includegraphics{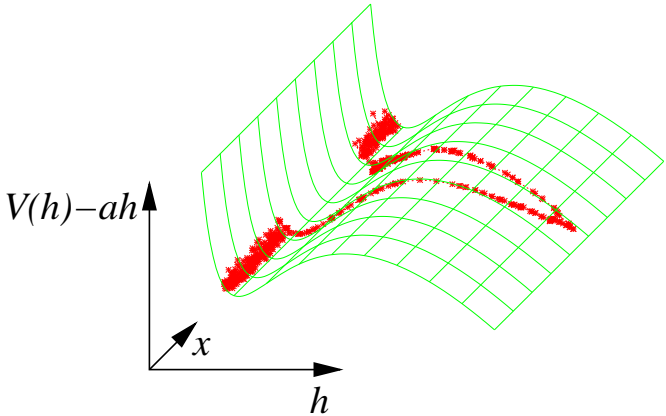}}
\caption{Typical potential and pinned interface, close to
the pinning-to-depinning transition line (rightmost transition in the
previous figure). Sites away from the potential-well are pulled by the
positive linear term $a > a^+$. In this way, characteristic triangular
fluctuations appear and are eliminated, in a region where the depinned 
phase is also stable.}
\label{triangle}
\end{center}
\end{figure}

On the other hand, the pinned phase loses its stability when the
density of pinned sites, i.e. sites within the attractive potential
well, vanishes. This happens at a value $a^*(b)~ > ~ a^+$ which
depends on $b$. It has been shown \cite{MNreview,synchro} that this
depinning transition is related to {\em directed percolation} (DP)
\cite{AS}. In figure \ref{triangle} we plot a typical pinned
interface within the coexistence region. Most of the sites lie within
the potential well. Those patches which, owing to fluctuations,
overcome the potential barrier are locally depinned and grow quickly 
to form triangular structures, with a well defined slope, anchored at the 
wall (figure \ref{triangle}). Driven by the
negative nonlinear term, $\lambda (\nabla h)^2$, these triangles
shrink at a constant velocity revealing the mechanism for the
elimination of islands of the depinned phase. Therefore, even if the
depinned phase is stable, initially pinned
interfaces will remain so, since a mechanism for the elimination
of islands of the depinned phase does exist. By contrast, for systems with
$\lambda > 0$, broad phase coexistence does not occur because triangular 
fluctuations are pulled away from the wall \cite{Haye,MNreview} and thus 
there is no mechanism to eliminate the minority phase islands.

As the stability threshold $a^*(b)$ is approached, the size of the 
depinned regions increases until they extend over the whole system. 
Then the triangular fluctuations cannot be eliminated and the interface 
is depinned. Consequently, for $b < 0$ pinned and depinned phases lose 
their stability at different values of 
the control parameter $a$. Between these values, there is a broad 
region of phase-coexistence where initially pinned interfaces remain 
pinned, while moving interfaces keep on moving. This remains the case
in the infinitely-large system-size limit \cite{finite}. 

Let us stress that while the transition at $a^*$ is controlled by
the effect of fluctuations that make pinned sites jump over the
potential well, the one at $a^+$ is controlled by the average velocity
of the free interface. Therefore two different mechanisms are at play.
{\it Essential for the broad phase-coexistence region is the 
asymmetric role played by the potential on pinned and depinned 
interfaces}: the potential has no effect on depinned interfaces, while it 
stabilizes pinned interfaces in a region where depinned interfaces are 
also stable. The potential acts as an asymmetric force that depends on the 
state of the interface. 

The dynamical asymmetry of the potential, however strong, cannot by itself 
produce generic bistability. 
A non-equilibrium ingredient is also needed, as can be seen by letting 
$\lambda=0$ in equation (\ref{kpzww}). This results in the equilibrium EW 
equation in the presence of an attractive potential, which describes the 
pinning-depinning transition of an equilibrium interface where no broad 
phase-coexistence exists. As in Toom's model, it is the combination 
of the nonequilibrium nature and the asymmetry of the dynamics that 
provides the mechanism for the elimination of minority phase islands, 
required for the existence of a broad region of phase coexistence.  

Before ending this section, let us remark that the phenomenology 
described above is not unique to the solutions of the continuum 
equation (\ref{kpzww}); it was first noticed in discrete models of 
nonequilibrium wetting \cite{Haye}. In addition, broad phase-coexistence 
is still observed when long-ranged potentials or slightly different 
nonequilibrium dynamics are considered.

\section{Pinning-depinning in disordered media}
\label{sec:3}

The third example to be discussed concerns the transitions between static 
and moving interfaces in the presence of quenched disorder. Relevant 
applications are ubiquitous and include
solid-on-solid friction experiments, charge density waves \cite{CDW},
wetting of rough surfaces \cite{Helium}, vortex lines in type II
superconductors \cite{HB}, and earthquakes \cite{terremotos}, to mention
but a few.

A familiar case is provided by elastic, extended objects sliding 
against rough surfaces. Segments of the sliding object may be pinned
by the inhomogeneities of the medium, only to be set into motion by
the elastic forces from neighboring regions that have overcome the
resistance of the medium and move freely. The final, average velocity
will depend on the strength of the applied external force. A
simplified model that purports to capture this phenomenology is
the  ``Quenched Edwards Wilkinson'' (QEW) equation (see 
\cite{Barabasi,HHZ})
\begin{equation}
\partial_t h({\bf x},t) = D \nabla^2 h ({\bf x},t) + F + 
\eta({\bf x},h({\bf x},t))~.
\label{lim}
\end{equation}
This equation describes an elastic interface, given by the height
profile $h({\bf x},t)$, with surface tension $D$, under the influence
of a constant external driving term $F$, and {\em quenched noise}
$\eta({\bf x},h({\bf x},t))$. It exhibits a continuous {\em
pinning-depinning} transition at a critical force $F_c$ from a pinned
phase to a moving one. The universal properties of this transition were
studied using renormalization group methods \cite{RG-LIM}, and the
critical exponents were measured both computationally and experimentally
\cite{Usadel,Rosso}.

The simplest way to simulate the dynamics of the universality class 
described by equation (\ref{lim}) is by using the Leschhorn cellular 
automaton \cite{Lesch}. A quenched random pinning force, $f(x,h(x))$, 
sampled homogeneously from the interval $[0,1]$ is assigned to each 
coordinate $(x,h(x))$, in one-dimension, and an interface profile 
$h(x,t=0)=1$ is taken as the initial condition. The dynamics then proceeds 
as 
follows: at every time step, and at every site $x$, a local force is given 
by the sum of {\bf i)} the discretized Laplacian at that site, 
($h(x+1)+h(x-1)-2h(x)$), and {\bf ii)} the constant force $F$. If the 
local force exceeds the pinning value, $f(x)$, then $h(x)$ is increased by 
one unit. The process is repeated for all sites and iterated in time. 
Below a given $F^*$ the system is pinned with probability one (in the 
infinite system-size limit), while for larger values of $F$ the interface 
advances with constant velocity. This phase transition is in the QEW 
universality class, as described by equation (\ref{lim}).

It is well known that the force required to start an object moving is 
greater than that necessary to keep it going. This is not captured by the 
QEW nor by the Leschhorn automaton due to a {\em 
no-passing} rule: an interface cannot overtake another one that is 
initially ahead of it. This is due to the fact that at every point of contact 
the pinning force is the same for both interfaces, whereas the elastic, 
restoring force on the advanced interface is greater than (or equal) 
the restoring force on the interface lagging behind. As a consequence of 
this rule, coexistence of moving and stationary interfaces is impossible 
since for a given external force the interface attains a unique velocity.

In an attempt to describing more realistic situations, {\it dynamic 
stress transfer mechanisms} were introduced recently in the Leschhorn 
automaton \cite{MS}. The local force at every site is increased by an 
extra contribution {\bf iii)} $M S(x)$, where $M$ is a control parameter,
and $S(x)$, a {\em stress-overshoot}, is equal to $1$ if either the
site at $x$ or at any of its nearest neighbors moved in the
preceeding time-step, and $0$ otherwise.  In this way, locally moving
interfaces are more likely to keep on moving, while the
stress-overshoots do not play any role on locally pinned
regions, where $S(x)=0$.  This will produce an effect similar to
that of inertia \cite{MS}. While the original Leschhorn model has 
a transition at $F^*$, the model endowed with stress-overshoots can
easily be seen to undergo a pinning transition at $F^{+}=F^*-M$. To see
this, note that at all moving sites the local force is increased by $M$ 
units, therefore the effective external force is
$F+M$ (see \cite{MS} for more details). On the other hand the
depinning transition is not affected by the stress-overshoots, and
therefore remains at $F^*$. This leads to a broad region of 
phase-coexistence delimited by $[F^{+},F^*]$. The transition at $F^+$ 
was reported to be continuous (in terms of velocities) and to belong to 
the QEW class for small $M$, and discontinuous above some value 
of $M$.
\begin{figure}
\begin{center}
\resizebox{0.5\textwidth}{!}{%
\includegraphics{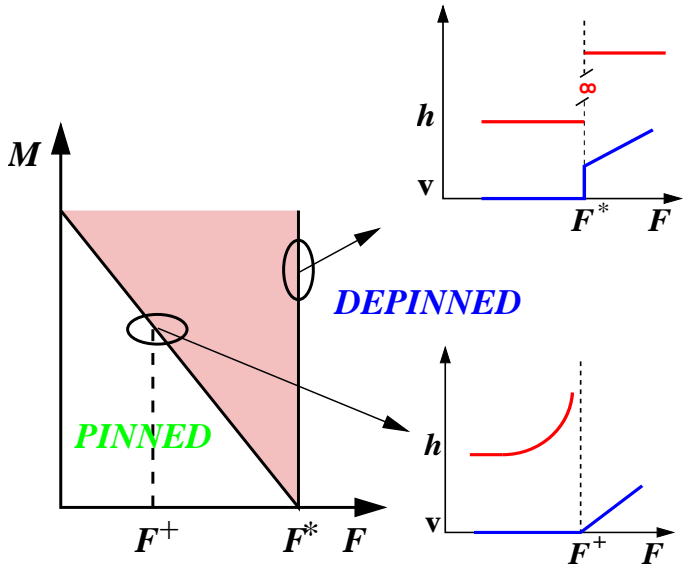}}
\caption
{Phase diagram as a function of $F$ and $M$. A broad phase-coexistence
region (shaded) exists. The order parameters (both average velocity
and average height) are plotted for both transitions delimiting the
broad coexistence region. In the lowermost (uppermost) panel the
height (velocity) changes discontinuously while the velocity (height)
changes continuously. The threshold in the $F^+$ branch, beyond
which the transition becomes discontinuous is not shown.}
\label{pd2}
\end{center}
\end{figure}
The origin of this broad region of phase-coexistence can be traced to
the fact that the stress-overshoots only affect the moving
phase. Therefore, regions expected to belong to the pinned phase in
the original model may become depinned due to the extra force
generated by inertia if they are initially moving, but remain pinned
if the interface is initially pinned. In a nutshell, {\it broad
coexistence results from the asymmetric role played by the
stress-overshoots in pinned and depinned interfaces.}  It should be
stressed that the hysteresis in this type of models is a {\it phony}
one, in the sense that it is destroyed by the inclusion of thermal
fluctuations \cite{MS}.

As in the previous section, asymmetry is only a necessary
condition, not a sufficient one; here, a robust mechanism for the
elimination of islands of the minority phase is also required. In this
case, it is essential that the inertial term depends on the state of
motion of the neighboring sites as well as on the site itself. The
mechanism, however, seems to be sensitive to the details of the model
\cite{MS,SF}.

Other universality classes in the interfaces-in-random-media realm are
the quenched KPZ equation \cite{Barabasi,HHZ,TKD}, the EW equation
with columnar noise \cite{Barabasi,columnar}, or the Mullins-Herring
equation \cite{Mullins}. All of them are susceptible of exhibiting broad
phase coexistence by an adequate inclusion of inertial effects.

\section{Discussion and conclusions}

Toom's NEC probabilistic automaton, introduced some 20 years ago, 
exhibits generic bistability. Its relation to recently proposed, 
nonequilibrium models should be noticed inasmuch as these 
models also exhibit broad phase-coexistence. 
In the latter, realistic driving forces, $a$ and $F$, favor 
interfacial motion, and are opposed by pinning forces, the 
attractive wall and the random medium, respectively. 
The latter depend crucially on the interfacial state.

The key feature of Toom's NEC model is the spatial 
{\em asymmetry of its dynamical rules}, of a type only possible in 
nonequilibrium systems, that results in an efficient {\em mechanism for 
the elimination of islands
of the minority phase}. Indeed, islands of minority spins shrink
after they have been created by fluctuations at a velocity roughly
independent of their radius. This is to be compared with the Ising model 
in zero magnetic field, where a droplet of radius $r$ shrinks with a
velocity proportional to $1/r$.

These two ingredients are also present, in more realistic terms, in
the interfacial models discussed in this paper, namely, a
nonequilibrium interface bound by an attractive wall (equation
(\ref{kpzww})), and an interface subject to stress-overshoots
advancing in a random medium (equation \ref{lim}). In particular,

{\bf i)} There is a clear asymmetry in both cases: the potential well
acts only on locally pinned regions, and inertia (stress-overshoots)
does so only on locally moving ones. Therefore, in one of the
coexisting phases the additional 'force' plays no role. This
dynamical asymmetry is the first essential ingredient for broad
phase-coexistence. Both the attractive potential and the
stress-overshoots have a clear physical origin, and play the same role
as the somewhat artificial spatial asymmetry of Toom's model.

{\bf ii)} A robust mechanism for the elimination of islands of the 
stable phase may be found in both examples. In the first case, droplets of 
the depinned phase, that could make the
interface detach from the wall, acquire a triangular shape and are
ultimately suppressed by the combination of the nonlinear force term 
$\lambda <0$ and the wall. The islands of depinned sites cannot act as 
nucleation bubbles and are eliminated in a time proportional to their 
size. Similarly, stress-overshoots foster the depinning of segments of the 
interface, when $F^+ ~ < ~ F ~ < ~F^*$, preventing nucleation of the 
pinned interface. As explained in the paper, modifications of these
nonequilibrium models that fail to provide such a mechanism, do not exhibit 
broad phase coexistence.

Note that while in the bound-interface model, it is the {\it pinned}
phase that is further stabilized by the asymmetric dynamics, in the 
stress-overshoots model it is the {\it depinned} phase that is further 
stabilized.

Another interesting issue concerns the nature and the universality 
class of the continuous transitions. Interestingly enough, for the 
model with stress-overshoots the continuous (in terms of $\langle v 
\rangle$) pinning transition, in the QEW universality class as in 
the original inertia-free Leschhorn automaton, persists for small $M$, 
whereas for larger $M$ it becomes discontinuous. On the other hand, 
despite previous claims, for the model of bound interfaces the depinning 
transition remains continuous (in terms of $ \langle h \rangle$), but the
universality class changes from the so-called {\it multiplicative
noise} class \cite{MNreview} for $b > 0$, to directed percolation for
$b < 0$ (attractive wall).

In conclusion, {\it the additional term required to generate broad 
phase-coexistence can be a relevant or an irrelevant perturbation to the 
continuous phase transition of the original model}.
Moreover, the order of the transition may or may not be affected by this 
term.

Finally, let us stress again, that in both interfacial models, one 
of the transitions delimiting the broad coexistence region can be 
continuous, at odds, with the intuition developed in equilibrium 
situations. 
Nevertheless, at least  one of the boundaries has to be discontinuous, 
as can be checked easily using continuity arguments (see figures 
\ref{pd1} and \ref{pd2}).

  {\bf Summing up}, two interfacial models exhibiting generic
  bistability have been discussed. Both are nonequilibrium
  models with an essential asymmetry in the dynamics. 
  The asymmetry is such that it eliminates islands of the minority 
  phase efficiently. The stability of one of the phases is 
  enhanced, resulting in generic bistability over a broad region of 
  parameter space. In both models this is achieved by including realistic 
  ingredients, namely an attractive wall and inertia. This puts generic 
  phase coexistence, with its many conceptual and applied consequences, 
  under a more solid and motivated physical basis.

\vspace{0.25cm}

{\small We acknowledge J. Schwarz for useful correspondence, as well as
S. Zapperi for useful comments and discussions. Support from the Spanish MCyT
(FEDER) under project BFM2001-2841 is also acknowledged.}


\begin{thebibliography}{99}
\bibitem{finite} 
In general, in finite systems that exhibit pinned (active) and
depinned phases (absorbing), depinned phases are the only stable ones
as fluctuations take the system to the absorbing state in a finite
time \cite{AS}.
\end{thebibliography}
\end{document}